\documentclass[12pt,preprint]{aastex}

\newcommand{\chandra}{{\it Chandra}}

\newcommand{\xlf}{XLF}
\newcommand{\xlfs}{XLFs}
\newcommand{\aciss}{ACIS-S}
\newcommand{\ergseight}{$\times 10^{38}$ erg s$^{-1}$}
\newcommand{\ergsseven}{$\times 10^{37}$ erg s$^{-1}$}
\newcommand{\lmxb}{LMXB}

\newcommand{\snr}{SNR}
\newcommand{\hmxb}{HMXB}

\shorttitle{The Luminosity Function of X-ray Sources in Spiral Galaxies}
\shortauthors{Prestwich et al.}

\begin{document}

\title{The Luminosity Function of X-ray Sources in Spiral Galaxies}

\author{A .H. Prestwich}
\affil{Harvard-Smithsonian Center for Astrophysics, Cambridge, MA 02138}
\author{R. E. Kilgard}
\affil{Wesleyan Univesity}
\author{F. Primini}
\affil{Harvard-Smithsonian Center for Astrophysics, Cambridge, MA 02138}
\author{J.C. McDowell}
\affil{Harvard-Smithsonian Center for Astrophysics, Cambridge, MA 02138}
\author{A. Zezas}
\affil{University of Crete,\\Harvard-Smithsonian Center for Astrophysics}

\begin{abstract}

X-ray sources in spiral galaxies can be approximately
classified into bulge and disk populations.    The bulge (or hard) sources have X-ray colors which are consistent with Low Mass X-ray Binaries (LMXB)  but the the disk sources have softer colors suggesting a different type of source.   In this paper, we
further study the properties of hard and soft sources by constructing
color segregated X-ray Luminosity Functions (\xlf) for these two populations.  Since the number of sources in any given galaxy is small, we 
coadded sources from a sample of nearby, face-on spiral galaxies
observed by Chandra as a Large Project in Cycle 2.    We use simulations to carefully correct the \xlf\ for completeness.   The composite  hard
source \xlf\  is not consistent with a single power-law fit.   At luminosities $L_x>$ 3\ergseight\ it is well fit by a power law with a slope that is  consistent with that found for sources in elliptical galaxies by 
\cite{Kim2004}.  This is supports the suggestion that the hard
sources are dominated by LMXBs.  In contrast, the high luminosity  \xlf\ of soft sources has a slope similar to the
``universal'' High Mass X-ray Binary \xlf.   Some of these sources are stellar mass black-hole binaries accreting at high rates in a thermal/steep power law state.   The softest sources have inferred disk temperatures that are considerably lower than found in galactic black holes binaries.   These sources are not well understood, but some may be super-soft ultra-luminous X-ray sources in a quiescent state as suggested by \cite{Soria2009}.

\end{abstract}

\section{Introduction}
\label{intro}
It has been known for three decades that nearby spiral galaxies
contain a multitude of X-ray sources \citep{van Speybroeck1979}.
Observations of the Milky Way and Local Group galaxies  have shown
that the brightest of these sources are Low Mass X-ray Binaries
(\lmxb).  In these systems, material is transfered at a high rate  via Roche Lobe
overflow from a low mass star onto a compact companion (a white dwarf,
neutron star or black hole).     Their X-ray spectra are broadly characterized by a power law with photon index $\gamma\sim$2.5 \citep{Tanaka1997}.   High Mass X-ray Binaries (\hmxb) are powered by Bondi-Hoyle accretion from the stellar winds of a young, high mass
star onto a compact object.     They typically have much harder spectra than \lmxb\  (photon index $\gamma$=0-1.5, \cite{Haberl2004})   The accretion rates are lower than for Roche Lobe overflow sources and the corresponding X-ray luminosities also  lower \citep{White1989}.    In addition to the
basic classification depending on the mass of the secondary, many
other groups of sources have been identified depending on their X-ray
spectra and variability properties.  For example, so-called Super Soft
Sources (SSS),  believed to be binaries with white dwarf primaries accreting matter at a highly super-Eddington rate  \citep{Greiner1991} have
 essentially all of their flux below 1keV, and are
highly variable \citep{Di Stefano2003, Di Stefano2004}.  Supernova remnants
are also known to be strong X-ray sources \citep{Reynolds2008}.    For a review of X-ray sources in galaxies  see \cite{Fabbiano2006}.

 In the closest galaxies, clues as to the nature of X-ray sources can be derived from the 
optical environments and/or counterparts \citep{Kilgard2005,Trudolyubov2005}.   However, it is more difficult to identify X-ray sources in galaxies which are too distant for optical counterparts to be detected and where there are not enough counts to get an X-ray spectrum.
In \cite{Prestwich2003}  we suggested that  X-ray colors are a good starting point for source classification.   We found that the X-ray colors of the sources described above (low and high mass binaries, SSS, and SNR) lie in different parts of the X-ray color-color diagram, albeit with some overlap of source types.   In particular,  sources in elliptical galaxies and spiral bulges fall in a
well defined region of the X-ray color-color diagram.  We suggest
that this  population is dominated by Low Mass X-ray Binaries.    We find very few sources with hard \hmxb\ colors,  which is to be expected  given their low X-ray luminosities.   We also find a  population of faint X-ray soft sources in disk galaxies not seen
in bulges.  These soft/thermal sources are clearly associated with
star formation, and are likely a mix of supernova remnants and accretion powered sources (see also \cite{Di Stefano2004}).

In this paper, we further investigate the properties of hard (or \lmxb) and
 soft
sources by constructing color segregated X-ray Luminosity Functions
(\xlfs).     Our motivation is to confirm our interpretation
that the hard sources are dominated by LMXBs and to better understand
the mix of objects that fall into the soft class.  In
Section~\ref{sec:s&s} we describe our galaxy sample and completeness
simulations and in Section~\ref{sec:fitting} we describe fits to hard and soft sources.   In Section~\ref{sec:pop} we compare our results to previous work and discuss the origin of soft sources.  Our conclusions are summarized in Section~\ref{sec:s&c}.

\section{Galaxy Sample and Simulations}\label{sec:s&s}

Most spiral galaxies have 10-50 X-ray sources, so that it is impossible to
construct  statistically significant color segregated luminosity
functions for a single galaxy.  We therefore constructed
color-segregated \xlfs\ by coadding hard and soft sources from 11
spiral galaxies.   The galaxies chosen for this work were observed as
a \chandra\  Large Project in Cycle 2 (see \cite{Kilgard2005} for details
of data reduction and a complete source list).    They are all nearby ($<$ 10 Mpc), 
face on spirals with low foreground  absorption.  The X-ray colors of 
sources in this sample are therefore relatively insensitive to patchy
local  absorption and the inclination of the galaxy.    We use color definitions hard color ($HC$) and soft color ($SC$) introduced by \cite{Prestwich2003}:

\[HC=\frac{H-M}{T}\]
\[SC=\frac{M-S}{T}\]

Here $H$, $M$,  and $S$ are counts in the soft (0.3-1 keV), medium (1-2 keV) and hard (2-8 keV) bands respectively  and $T=S+M+H$.     We define soft and hard source colors as follows:

\begin{tabular}{ll}
soft color: &  $SC < 0.4$ \\
hard color: & $-0.4< SC<0.2, -0.3 < HC <0.4$\\
\end{tabular}

\begin{figure}
 \plotone{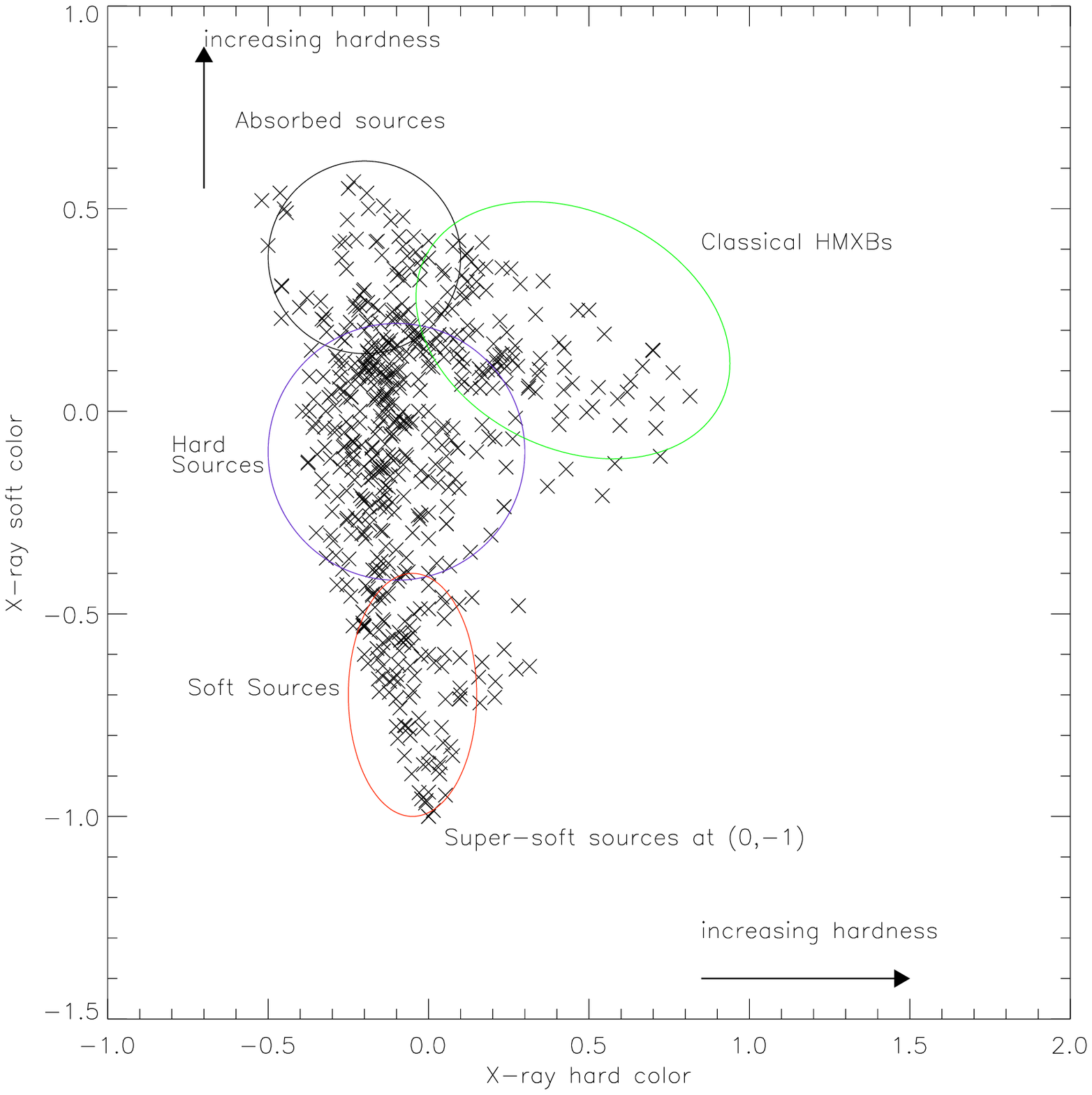}
\caption{ X-ray color-color diagram for sources in spiral galaxies from \ref{Prestwich2003}.  The location of hard and soft sources is shown}
\label{xcols}
\end{figure}

These regions are shown on a color-color diagram in Figure~\ref{xcols}.

\begin{figure}
 \plotone{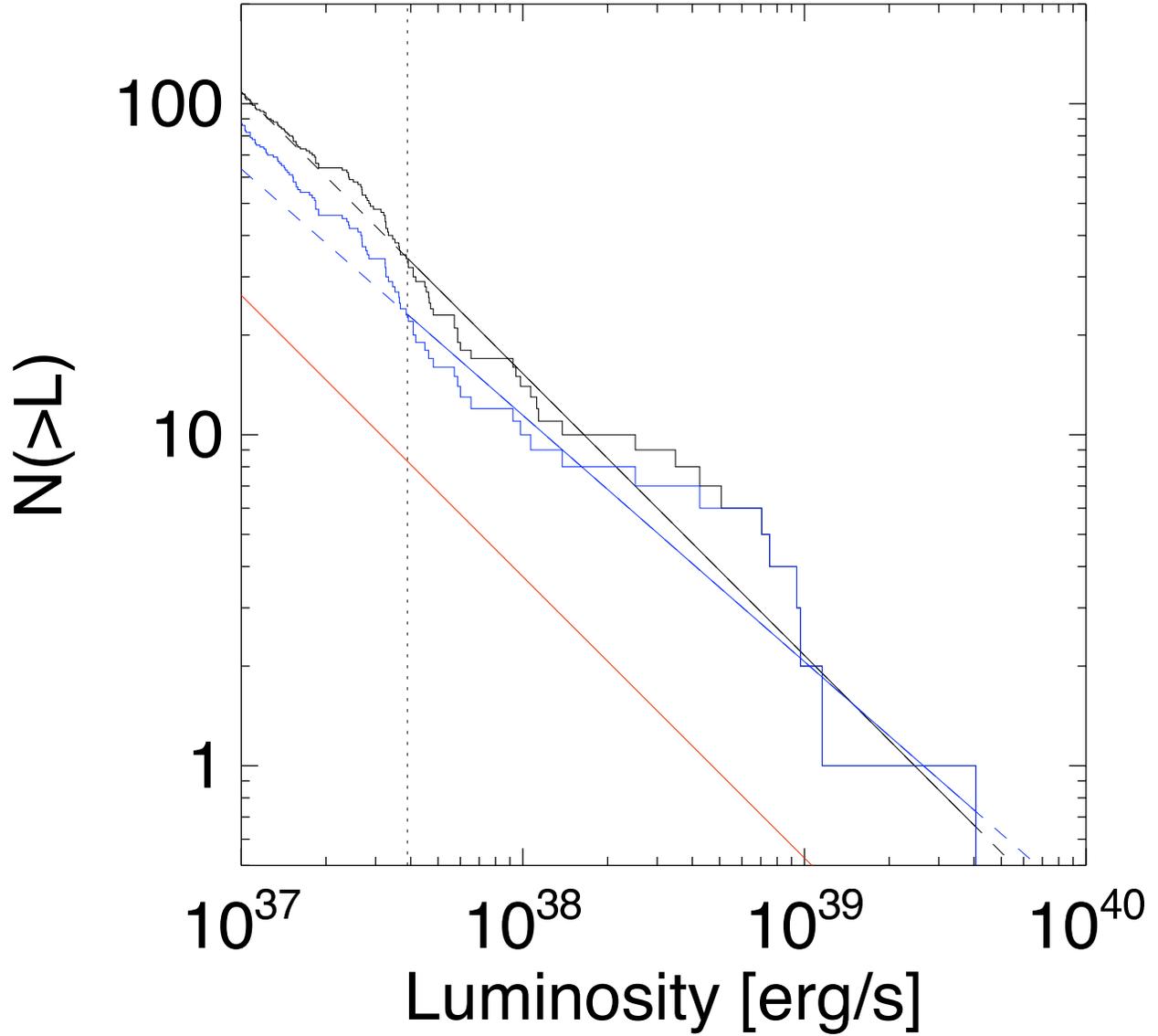}
\caption{\xlf\  for soft sources.   The black curves show the data and best fit power law, the red  line shows  the distribution of background sources and the blue curves the background corrected \xlf.  The vertical dotted line shows the completeness limit.   Points below the completeness limit were not included in the fit.}
\label{soft_xlf}
\end{figure}

\begin{figure}
 \plotone{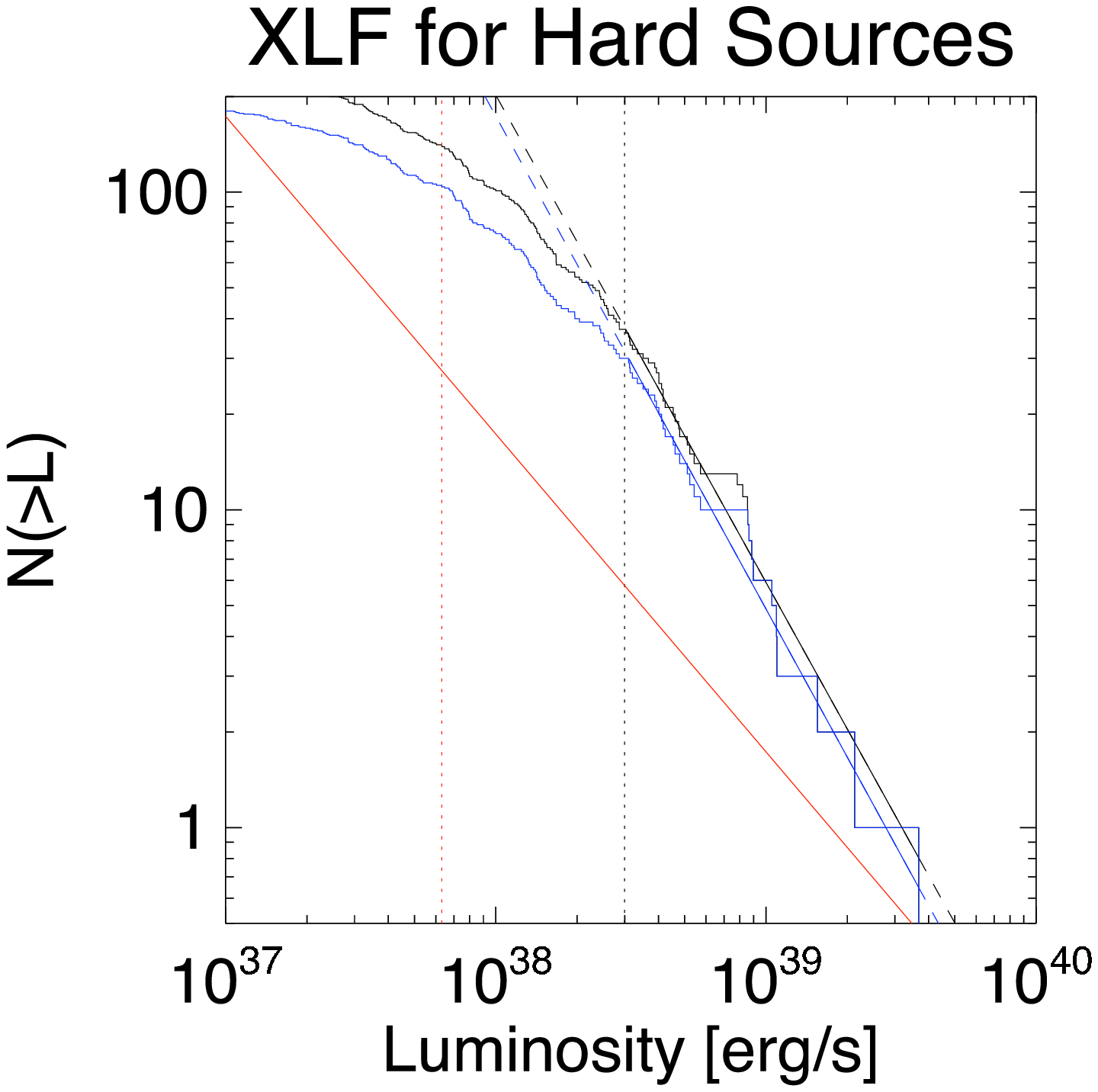}
\caption{\xlf\ for hard sources.   The black curves show the data and best fit power law, the red  line shows  the distribution of background sources and the blue curves the background corrected \xlf.  The red vertical dotted line shows the completeness limit.   The black vertical line shows the lowest luminosity included in the fit. }
\label{hard_xlf}
\end{figure}

The composite \xlfs\ for soft and hard sources sources
are shown in Figures ~\ref{soft_xlf} and \ref{hard_xlf}. Inspection of these figures suggest
that the soft sources are well represented by a single power law. In contrast, the \xlf\ of the
hard sources starts to flatten at $L_x \sim3$\ergseight.  The
flattening of the \xlf\ may be due to a real change in the properties
of the sources or an artifact due to incompleteness (or a combination
of both).   There are several
factors that affect the detection threshold for X-ray sources,
including the presence of diffuse emission and the distance of the
source from the aimpoint. In addition, \aciss\ was significantly more sensitive to 
soft emission ($<$ 1 keV) in Cycle 2 than to harder sources; this will 
result in a different completeness limit for hard and soft sources.
We construct simulated \xlf\ to help distinguish between a real break in the
\xlf\ and incompletemess effects.

We use the ``backward'' method described by \cite{Kim2004} to ascertain the 
completeness limit of each galaxy for a given source type (hard or
soft).  This method involves simulating a source with the appropriate
spectrum, and adding the simulated source to the real galaxy data.
 The source detection algorithm used for the original galaxy source
list is then run again to determine whether the ``fake'' source is
detected.  Approximately 20,000 soft sources and 20,000 hard sources
are added to each galaxy, sprinkled randomly over the  area within  $D_{25}$ .    We use a power law of photon index 1.2 to simulate hard sources and a power law of photon index 3.0 to simulate soft sources.        The detect algorithm was run on the full band image, as was also the case for the galaxy analysis.  The ratio of the number of input to the
number of detected sources at a given luminosity is the key factor:
when this falls below 100\% the sample is not complete.   In order to
construct a composite \xlf, we added sources to
luminosity bins where the completness fell below 100\%.  For each
galaxy, sources were added to the \xlf\ according to the following formula:

\begin{equation}
N_{add}(L)=\frac{(1-F)}{F}N_{gal}(L)
\end{equation}

Here  $N_{add}(L)$ is the number of sources added to the  bin of luminosity
$L$ and   $F$ is the fractional completeness, defined as the ratio of the
number of input to the number of sources detected in simulations.  $N_{gal}(L)$ is the number of real sources in the galaxy with luminosity $L$.   

We find that the soft source population is 100\% complete down to
3.9\ergsseven\  for most galaxies (M101, IC 5332, M94, M51, M83, NGC 2681, M74 and NGC 1291).  No sources were added to the \xlf\ for these
galaxies.  We omit galaxies NGC 4314, NGC 3184 and NGC 278 from the soft
source composite \xlf\ because they are less than 40\% complete at
3.9\ergsseven.    Detection of hard sources was complete down to
7.5\ergsseven\ for galaxies IC 5332, M101, M51, M83, NGC 2681, NGC 278
and NGC 628.  Sources were added to the lowest luminosity bins of the
remaining galaxies as
summarized in Table~\ref{hard_comp}.  In addition to correcting for
completeness in the lowest luminosity bins, we also removed bright nuclear
sources from the complosite \xlfs.  In many cases these are weak
LINERS and not X-ray binaries.

\section{Fits to Composite \xlf}\label{sec:fitting}

The method used for fitting is described by \cite{Kilgard2005} and full details can be found in that paper.    Here we give a brief summary of the procedure.

We fit a single power law to the unbinned, differential, composite 
XLFs  using a maximum likelihood statistic \citep{Crawford1970}.    The \xlfs\ were corrected for completeness as described in the previous section.  The single power law has the functional form:

\begin{equation}
 \frac{dN}{dL_{X}}\sim L_{X}^{-\gamma}
\end{equation}

The goodness-of-fit estimate is performed by simulating a luminosity distribution with the best fit slope.   One million iterations were performed.  If the data are well fit by a single power law the goodness-of-fit statistic (GOF) approaches 1.0.    The contribution of cosmic background sources to the \xlf\ was evaluated using  the LogN-LogS curves of \cite{Giacconi2001}.  The soft band was used for the soft sources and the hard band for the hard sources.    We fit down to the completeness limit (3.9\ergsseven) for soft sources.  It is impossible to fit a single power law from the completeness limit to the highest luminosity source for the hard sources.   We therefore fit a single power  law above  3\ergseight.    Figures \ref{soft_xlf} and \ref{hard_xlf}  show the composite \xlf\  and best fit power law (black curves)  for soft and hard sources.  Also shown is the slope representing the background source distribution (red curve) and the background subtracted data and best fit (blue curve).  Best fit parameters to the composite  \xlf\  are shown in Table~\ref{tb:fits}.

\section{The X-ray Source Population in Spiral Galaxies}
\label{sec:pop}

There are two basic conclusions to be drawn from the results presented in the previous section.  The first is that although the completeness limit for a given observation depends on many factors, such as diffuse emission,  it is lower for soft sources than for hard sources.   The second is that there appears to be a break in the hard source  \xlf\ at about $\sim$3\ergseight.     The flattening of the  hard source \xlf\ shown in Figure~\ref{hard_xlf} at low luminosities ($<$7\ergsseven) is largely due to incompleteness.     
\subsection{Hard Sources}

In \cite{Prestwich2003} we suggest that the hard sources are dominated by Low Mass X-ray Binaries.   The \xlf\ of sources in elliptical galaxies, which are almost exclusively \lmxb\, has been studied in depth by \cite{Kim2004} and \cite{Gilfanov2004}.    \cite{Kim2004} find that the \xlf\ of a sample of 14 elliptical galaxies (corrected for incompleteness) can be fit with a single power law, $\gamma$=1.9 over the range 2$\times10^{37}-2\times10^{39}$ ergs s$^{-1}$.   Although not formally required, a broken power law gives a better fit ($\gamma_{1}$=1.8$\pm$0.2, $\gamma_{2}$=2.8$\pm$0.6) with a break luminosity L$_{b}$=5$\times10^{38}$ ergs s$^{-1}$.    Our fits to the hard sources show very similar features: an \xlf\ which steepens at a few $\times10^{38}$ ergs s$^{-1}$.   Our fit parameters are consistent with those of  \cite{Kim2004}, albeit with large uncertainties due to a smaller sample.  It is likely that we are sampling the same population as  \cite{Kim2004}.

\subsection{Soft Sources}

 In \cite{Prestwich2003} we suggested that many of the soft sources are thermal supernova remnants (\snr).    We stress here that most of the soft sources considered in \cite{Prestwich2003} have luminosities $< $1.0\ergsseven.  They are below the  completeness limit for the composite \xlf\ derived here.    The soft sources at the high luminosity end of the \xlf\  are almost certainly dominated by  accretion powered sources.     The association of soft sources with star formation strongly suggests that the donor stars are young supergiant, O or B stars, even though the soft colors of these binaries is very different from the hard spectra of wind-accretion sources described in Section \ref{intro}.     We therefore compare the \xlf\ of soft sources to that of \hmxb.   The nature of the soft sources is discussed further in Section~\ref{sec:soft}
 
 The \xlf\ of high mass binaries in star forming  galaxies has been studied in detail by \cite{Grimm2003}.     They find that the \hmxb\ \xlf\ is well described by a single power law ($\gamma=1.61\pm0.12$) over the luminosity range $10^{35}-10^{40}$  ergs s$^{-1}$.    The slope derived by \cite{Grimm2003} is very close to our best fit single power law for soft sources ($\gamma=1.73\pm0.15$), suggesting that we are sampling the same population.

 \section{The Nature of the Soft Sources}
 \label{sec:soft}
 
 As discussed above, the luminous soft sources discussed here are most likely \hmxb.  However, they are clearly more luminous and have much softer spectra than wind-fed \hmxb\ (binary pulsars) found in the Milky Way and local group galaxies.    The higher luminosities suggest that the soft sources are accreting material at a higher rate than is found in wind-fed HMXB.    This requirement, plus the softer spectra, suggests that the accretion mechanism is fundamentally different than galactic \hmxb.   In this section we discuss the nature of these sources in more detail.
 
\begin{figure}
 \plotone{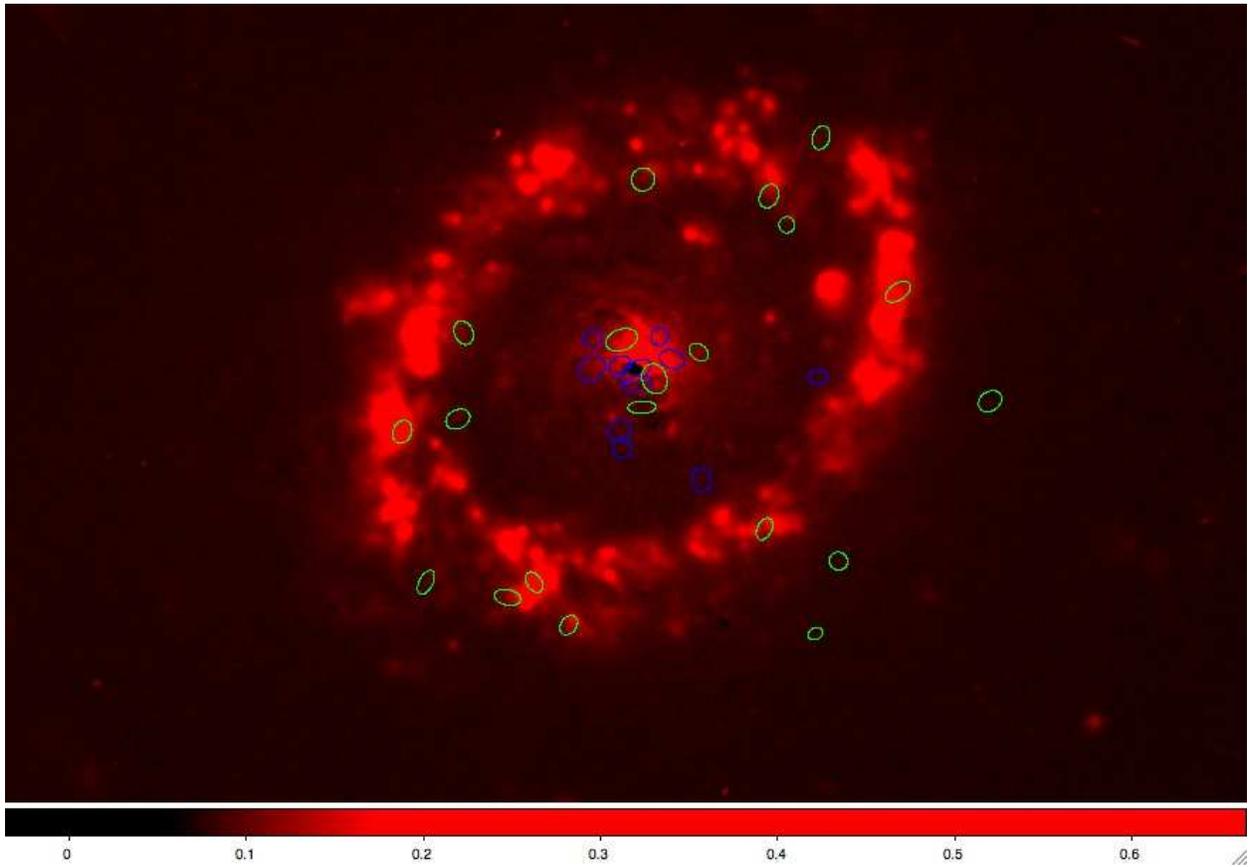}
\caption{ H$\alpha$ image of the center of NGC 4736 showing the star forming ring.  Positions of soft X-ray sources are plotted in green and hard sources plotted in blue.  The soft sources are predominantly associated with the ring.  }
\label{fg:n4736}
\end{figure}

\begin{figure}
 \plotone{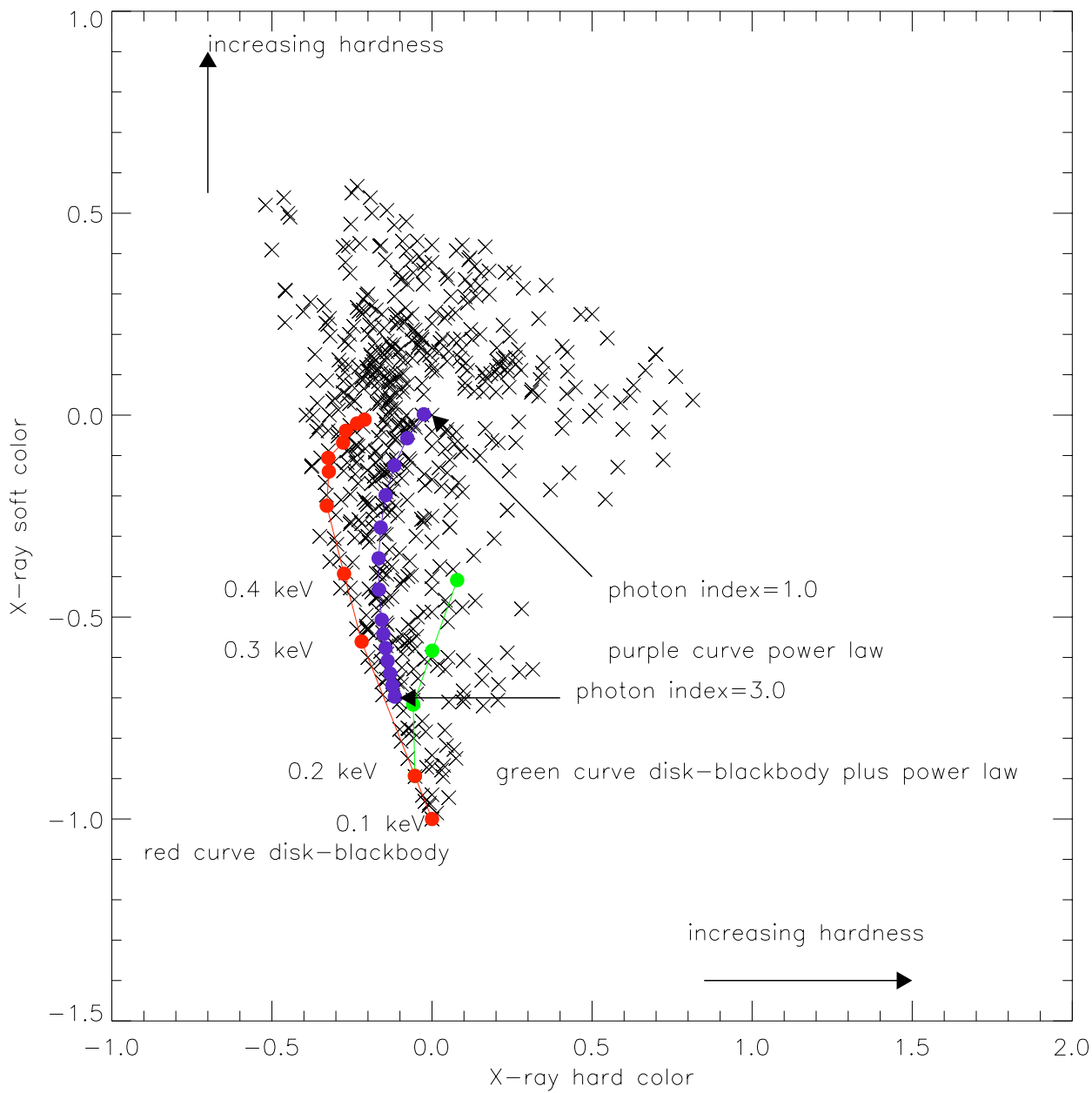}
\caption{ X-ray color-color diagram for sources in spiral galaxies.    Curves for thermal and SPL models for black binaries are plotted for reference.  The red curve shows the colors of disk-blackbody models for inner disk temperatures of 0.1-1.1 keV.       Adding a power law component to a disk-blackbody has the effect of moving the colors up and to the left of the diagram, as demonstrated by the green points.   The purple points show a simple power law characteristic of sources in the SPL state..  
 }
\label{xcols_bh}
\end{figure}

  The association of the soft sources with star formation is clearly demonstrated in the case of NGC 4736 (M94).  This galaxy has a circumnuclear ring of star formation (\cite{Waller2001}).   Figure~\ref{fg:n4736} shows the H$\alpha$ image of this galaxy, with the positions of soft (green) and hard (blue) X-ray sources.   It is clear that the soft sources are found primarily in the star forming ring.   Given the association with star formation,  these sources are most plausibly neutron star or black hole X-ray binaries with young stellar companions.   The X-ray spectral states of black hole X-ray binaries have been described in detail by \cite{Remillard2006}.    In the low accretion state the X-ray emission is dominated by a hot corona.  As the accretion rate increases the X-ray flux comes primarily from a disk and is thermal in origin.  The emission is characterized by a disk-blackbody spectrum (kT$\sim$1 keV) with some contribution at higher energies from a power law component.   In the highest accretion state (Steep Power Law, or SPL)  the X-ray emission is characterized by a soft power law ($\Gamma >$2.5).      Figure~\ref{xcols_bh}  shows the X-ray color-color diagram from \ref{Prestwich2003} for all detected sources, with thermal and SPL models  plotted for reference.   The red curve shows the colors of disk-blackbody models for inner disk temperatures of 0.1-1.1 keV.       Adding a power law component to a disk-blackbody has the effect of moving the colors up and to the left of the diagram, as demonstrated by the green points.   The purple points show a simple power law, representing the SPL state.   
  
 Figure~\ref{xcols_bh} shows that some of the soft sources have colors that are very similar to black holes accreting in the thermal or SPL state.  However, some of the softest have disk temperatures that are much cooler (kT$\sim$0.1-0.3 keV) than is typically found in galactic black hole binaries (kT$\sim$1 keV).     Low disk temperatures have also been observed in Ultra-Luminous X-ray Sources (ULX). Since $M\sim T_{in}^{-2}$, such low disk temperatures have been used to suggest that ULX contain Intermediate Mass Black Holes (IMBH, e.g. \cite{Makishima2000}).   It seems very unlikely that we observing a population of IMBH.  If the mass of the black holes are $\sim 100$ $M_{\odot}$ , the low luminosities ($10^{37}-10^{38}$ ergs s$^{-1}$) would imply sub-Eddington accretion rates.   In which case, we would not expect to see thermal emission, which is generally observed when the accretion  is dominated by a disk and the rates are close to Eddington.   The combination soft spectrum and low luminosity displayed by a subset of our soft sources is hard to explain if the sources are IMBH.   If they are stellar mass black holes, their disk  temperatures are unusually low.
 
 \cite{Soria2009} recently discussed a state transition in a ULX in NGC 4631.   In the high state NGC 4631 is a super-soft source.   In a low state observed by \chandra\ it has a luminosity $L_x\sim10^{37}-10^{38}$ ergs s$^{-1}$ and colors consistent with a thermal temperature of 0.1-0.3 keV.    They interpret the ULX/super-soft phase of NGC 4631 as a super-Eddingon outburst powered by nuclear burning on the surface of a white dwarf.   In it's low state, the  NGC 4631 ULX is very similar to the soft sources described here.   By extension, some soft sources in spiral galaxies may be quiescent super-soft ULX.   A detailed study of the variability properties of these sources is required to better understand their nature.

 \section{Summary and Conclusions}
 \label{sec:s&c}
 In this paper, we construct color  segregated X-ray Luminosity Functions (\xlf) of hard and soft
sources by combining data from a sample of face-on spiral galaxies.     Since the number of sources in any given galaxy is small, we 
coadded sources from a sample of nearby, face-on spiral galaxies.   To determine the compleness level of each galaxy as a function of source type (hard or soft)  we add  "fake" sources to real galaxy data and determine the ratio of fake sources added to the number recovered by source detection routines.    If all fake sources were detected the galaxy is 100\% at that flux level.    We added sources to the lowest luminosity bins to produce composite color segregated \xlf\ that are complete down to 3.9\ergsseven\ for soft sources and 7.5\ergsseven\ for hard sources.     

The composite hard
source \xlf\  is best fit by a broken power law and has a high luminosity
slope consistent with that found for sources in elliptical galaxies by 
\cite{Kim2004}.     We conclude the hard population is dominated by \lmxb.  The soft \xlf\  above the completeness limit is very similar to the "universal" \hmxb\ \xlf\  described by \cite{Grimm2003}.  This strongly suggests that although thermal supernova remnants  contribute to the soft population at lower luminosities,  the high luminosity end of the \xlf\ is dominated by accretion-powered \hmxb.    The association of soft sources with star formation is confirmed by looking in detail at NGC 4736, where the soft sources are coincident with a ring of H$\alpha$ emission.    Some of these sources are stellar-mass black hole binaries accreting in the thermal/PL states.   However, the  softest sources have inferred disk temperatures that are considerably lower than those found in galactic black holes binaries.   These sources are not well understood, but some may be super-soft ultra-luminous X-ray sources in a quiescent state as suggested by \cite{Soria2009}.

\section{Acknowledgments}

  This work was supported by NASA
contract NAS 8-39073 (CXC) and GO1-2029A.  Thanks to an anonymous referee for several very useful suggestions.

\begin{deluxetable}{ccccc}
\label{hard_comp}
\tablecaption{Completness Limits for Hard Sources \label{hard_comp}}
\tablewidth{0pt}
\tablehead{
  \colhead{Luminosity} & \colhead{NGC 4314} & \colhead{NGC 1291} &
\colhead{NGC 3184}  & \colhead{M94}\\
\colhead{\ergseight} & & & & \\
}
\startdata
0.75 & 20\% (5) & 50\% (4) &50\% (1) & 70\% (1)\\
0.95 & 50\% (2) & 68\% (1) &70\% (1) & 90\% (1)\\
1.15 & 60\% (1) & 77\% (1) &80\% (0) & 100\% (0)\\
1.35 & 80\% (0) & 90\% (0) &90\% (1) & 100\% (0)\\
1.5  & 90\% (1) & 100\% (0)&100\% (0)& 100\% (0)\\
\enddata
\tablecomments{Table gives completeness as a percentage of input
sources detected.  The number in parentheses is the number of sources
added to the \xlf\ to account for completness. Galaxies not listed are
complete down to 7.5\ergsseven.} 
\end{deluxetable}

\begin{deluxetable}{ccccccccccccc}
\tabletypesize{\scriptsize}
\tablecaption{Power law fits to composite, corrected \xlf}
\tablewidth{0pt}
\tablehead{
\colhead{Sample} 
& \colhead{$\gamma$}
& \colhead{$\gamma_{corr}$}
&\colhead{L$_{low}$}
  & \colhead{L$_{comp}$}
  &\colhead{$GOF$} 
 & \colhead{$N_{fit}$} 
 & \colhead{$N_{comp}$} 
 & \colhead{$N_{tot}$} 
 \\
  & &
  &\colhead{10$^{37}$ erg s$^{-1}$ } 
  &\colhead{10$^{37}$ erg s$^{-1}$ }
   & & &  \\
}
\startdata
Soft 
& 1.82$\pm$0.14 
& 1.73$\pm$0.15
& 3.9
& 3.9
& 0.7
& 32 
&32
&170
\\
Hard 
 & 2.53$\pm$0.25
 &2.55$\pm$0.28
  & 30.0
  & 7.5 
  & 1.05
  & 37 
  &104
  &236
  \\
\enddata
\tablecomments{Power law fits to composite, corrected \xlf.  $\gamma$ is the fitted slope to the \xlf\ with no background correction, $\gamma_{corr}$ correcting for background sources.   $L_{low}$ is the low luminosity used in the fit, $L_{comp}$ the completeness limit.   GOF is the goodness-of-fit parameter and $N_{fit}$ the number of sources included in the fit.   $N_{comp}$ is the number of sources above the completeness limit and $N_{tot}$ is the total number of detected sources.} 
\label{tb:fits}
\end{deluxetable}

\end{document}